\documentclass[pra,showpacs,floatfix,%superscriptaddress,
groupedaddress,
longbibliography,
twocolumn]{revtex4-1}
\usepackage{graphicx}
\usepackage{epstopdf}
\usepackage{amssymb}
\usepackage{amsmath}
\usepackage{dsfont}
\usepackage{xspace}
\usepackage{color}
\usepackage{ulem}
\usepackage{xcolor}
\usepackage{hyperref}
\usepackage{subfigure}

\newcommand{\mean}[1]{\left\langle #1 \right\rangle}

\newcommand{\ket}[1]{\left\vert #1 \right\rangle}
\newcommand{\up}{\ket{\uparrow}}
\newcommand{\down}{\ket{\downarrow}}
\newcommand{\carb}{$^{13}\mbox{C}$~}
\newcommand{\NVm}{NV$^-$}

\newcommand{\mbf}[1]{\mathbf{#1}}

\begin{document}

\title{Decoherence-free quantum register of nuclear spins in diamond.}
\author{Francisco J. Gonz\'alez}
\affiliation{Centro de Investigaci\'on DAiTA Lab, Facultad de Estudios Interdisciplinarios, Universidad Mayor, Chile}
%\author{Jer\'onimo R. Maze}
%\affiliation{Instituto de F\'{\i}sica, Pontificia Universidad Cat\'{o}lica de Chile,
%Casilla 306, Santiago, Chile}
\author{Ra\'ul Coto}
\email{raul.coto@umayor.cl}
\affiliation{Centro de Investigaci\'on DAiTA Lab, Facultad de Estudios Interdisciplinarios, Universidad Mayor, Chile}
\date{\today}

\begin{abstract}
Solid-state quantum registers are exceptional for storing quantum information at room temperature with long coherence time. Nevertheless, practical applications toward quantum supremacy require even longer coherence time to allow for more complex algorithms. In this work we propose a quantum register that lies in a decoherence-free subspace to be implemented with color centers in diamond. The quantum information is encoded in two logical states composed of two nearby nuclear spins, while an electron spin is used as ancilla for initialization and control. Moreover, by tuning an off-axis magnetic field we enable non-nuclear-spin-preserving transitions that we use for preparing the register through Stimulating Raman Adiabatic Passage. Furthermore, we use this sequence to manipulate the quantum register and an individual nuclear spin.    
\end{abstract}

\maketitle

\section{Introduction}

Color centers in diamond are remarkable systems for quantum sensing \cite{Degen_2017,Balasubramanian_2008} and quantum information processing \cite{Wrachtrup_2006,Hensen_2015,Atature2018,Awschalom2018}. In particular, the negatively charged Nitrogen Vacancy center (\NVm)  provides a local electron spin with good coherence time, that can be initialized, control and readout with high fidelities \cite{Balasubramanian2009,Bar-Gill,Fuchs,Blok,Jiang09, Robledo2011}. Furthermore, there is a set of nearby nuclear spins that can be controlled through the \NVm via hyperfine coupling \cite{Zhang19,Hedge20}, allowing the creation of a hybrid electron-nuclear spins platform. This composite spin system enables different functionalities like error-correcting protocols \cite{Waldherr2014,Taminiau2014}, preparation of long-lived entangled states \cite{Neumann1326,Pfaff2013,Bradley} and quantum registers \cite{Bradley,Dutt07,Shim13}. Nuclear spins have a small gyromagnetic ratio that shields them from the environment, and thus, can be used as quantum memories thanks to longer coherence time. However, they have a slow response to control fields. Therefore, for practical applications we require extended coherence times plus fast control gates. In order to increase the coherence time further, one approach consist in preparing the system in a decoherence-free subspace \cite{Reiserer2016,Rama2020}. Hence, quantum registers and entangled states are passively protected from the environment, in contrast to active techniques like dynamical decoupling \cite{vanderSar,Bradley} or error-correction~\cite{Layden20}. Moreover, one can design pulse sequences that take advantage of the hyperfine coupling to effectively perform faster one- and two-qubits gates.   

In this work, we propose the realization of a decoherence-free quantum register composed of two nuclear spins. We show that the register is robust against spins transverse relaxation for both independent and common reservoir. The preparation of the register is achieved by properly tuning an off-axis magnetic field. The slow nuclear spin control through radiofrequency is replaced by an all-microwave control that exploits the hyperfine coupling to the \NVm, to individually and collectively address the nuclear spins \cite{Coto2017,Huillery}. The population transfer between relevant states follows non-nuclear-spin-preserving transitions allowed by the perpendicular component of the magnetic field. In order to improve selectively and gate fidelity, we use Stimulated Raman Adiabatic Passage (STIRAP)~\cite{Gaubatz90,Bergmann,Bergmann15,Vitanov}  to control the nuclear spins in the microwave regime. The STIRAP sequence adiabatically transfers population through a dark state and it has been demonstrated, theoretically \cite{Coto2017} and experimentally \cite{Zhou17}, in \NVm centers.

The reminder of this paper is structured as follows. Section \ref{section_spin} introduces the composite spin system. Section \ref{section_decohe} describes our quantum register encoded in the decoherence-free subspace, its performance against noise, preparation and control.  Section \ref{section_single} describes the control of an individual nuclear spin, and Section \ref{section_conclu} provides the final remarks of this work.

\section{Multi-spins system}\label{section_spin}

\begin{figure}[t]
\centering
\includegraphics[scale=0.22]{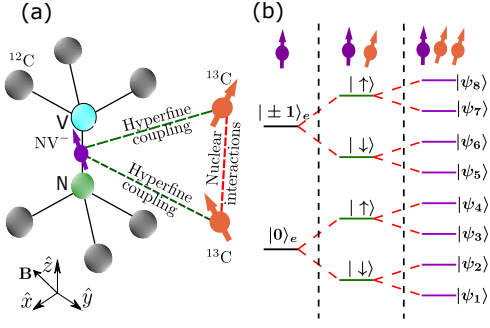}
\caption{(a) Tri-partite system given by an \NVm electron spin and two proximal \carb nuclear spins. (b) Energy levels of the tri-partite system. }
\label{fig1}
\end{figure}

Multi-spins systems provide a powerful platform to implement quantum registers \cite{Dutt07,Jiang09,Bradley,Waldherr2014,Taminiau2014}. In particular, this theoretical proposal is implemented with an electron spin ($S=1$) of a negatively charged Nitrogen Vacancy center (\NVm) in a natural diamond sample, with $1.1\%$ natural abundance of Carbon-13 (\carb) nuclear spins ($I=1/2$) \cite{Awschalom2018}. The \NVm ground state, labeled by $m_s=0$ and $m_s=\pm 1$, has a degeneracy in the magnetic sublevels $m_s=\pm 1$, which can be lifted by the application of an external magnetic field oriented in the \NVm quantization axis (z-axis). For the \carb nuclear spin, magnetic sublevels are labelled as $m_I=1/2$~($\up$) and $m_I=-1/2$~($\down$). The \NVm interacts with nearby $^{13}\mbox{C}$ via hyperfine coupling,  $\mathbf{S}\cdot\sum_{i}\mathbb{A}^{(i)}\cdot\mathbf{I}_i$, being $\mathbb{A}^{(i)}$ the hyperfine coupling tensor for the $i$th nuclear spin with components $A_{ij} = A_c\delta_{ij} + A_d(\delta_{ij} - 3\hat{r}_i\hat{r}_j)$. $A_c$ stands for the isotropic Fermi contact term and $A_d$ for the  dipolar interaction. Nuclear spins interact via dipolar coupling, with Hamiltonian $H_{nn}$ \cite{Dutt07,Jiang09} 
\begin{equation}\label{dipolar}
H_{nn}=\sum_{i<i}\frac{\mu_{0}\gamma_{c}^{2}}{4\pi r_{ij}^{3}}\left(\mathbf{I}_i\cdot \mathbf{I}_j-\frac{3(\mathbf{I}_i\cdot \mathbf{r}_{ij})(\mathbf{r}_{ij}\cdot \mathbf{I}_j)}{r_{ij}^2}\right),
\end{equation}
with $\mu_{0}$ the vacuum permeability, $r_{ij}$ is the distance between the $i$th and $j$th nuclear spins, and $\gamma_c$ is the gyromagnetic ratio of the nuclear spins. In Fig.~\ref{fig1} (a) we illustrate  these interactions for a tri-partite system given by the \NVm and two \carb nuclear spins, while in Fig.~\ref{fig1} (b) we show the energy levels for the spins configuration.

 The Hamiltonian considering two nuclear spins reads ($\hbar=1$),
\begin{eqnarray}\label{hamilt}
 H &=& DS_{z}^{2} + \gamma_e\mbf{B}\cdot\mbf{S} + \mbf{S}\cdot\sum_{i=1}^2\mathbb{A}^{(i)}\cdot\mbf{I}^{(i)}+ \gamma_c\mbf{B}\cdot\sum_{i=1}^2\mbf{I}^{(i)} \nonumber \\
 	&+& H_{nn},
 \end{eqnarray}
where $D/2\pi = 2.87$ GHz is the zero-field splitting of the NV$^{-}$, $\gamma_e/2\pi \approx 2.8$ MHz/G,  and $\gamma_c/2\pi \approx 1.07$ kHz/G are the gyromagnetic ratios of the electron and \carb nuclear spins, respectively. Without loss of generality, we consider the external magnetic field $\mbf{B}=B_{z}\hat{z} + B_{x}\hat{x}$. The $B_{z}$ component lifts the degeneracy, which allows to address a two-level manifold, say $\lbrace m_s=0,m_s=+1\rbrace$. The $B_{x}$ component enables electron-nuclear spin transitions that, as we shall see later on, are otherwise forbidden.

Considering the zero field splitting to be larger than the perpendicular magnetic field and the hyperfine coupling, i.e. $D\gg \gamma_eB_x$ and $D\gg A_{ij}$, one can perform the secular approximation, that neglects $S_x$ and $S_y$ contributions in the second and third terms in the Hamiltonian (\ref{hamilt}). Hence, the Hamiltonian for the tri-partite system can be written conditioned to the electronic spin manifold, such as
 \begin{align}
 H^{m_s}&=(m_s^2 D+m_s\gamma_eB_z)+\gamma_c\mathbf{B}\cdot(\mbf{I}^{(1)}+\mbf{I}^{(2)})\nonumber\\
 &\quad+m_{s}\sum_{n=x,y,z}\mathbb{A}_{z}^{(1)}\hat{I}_{n}^{(1)}+m_{s}\sum_{n=x,y,z}\mathbb{A}_{z}^{(2)}\hat{I}_{n}^{(2)}\nonumber\\
 &\quad+H_{nn}.
 \end{align}

For simplicity, we consider $\mbf{r}_{ij}$ oriented along the quantization axis ($\mbf{r}_{ij}\parallel \hat{z}$), that yields the following Hamiltonian in the $m_s=0$ manifold
 \begin{align}\label{hamilms_0}
 H^{m_s=0}&=\gamma_cB_{z}(\hat{I}_{z}^{(1)}+\hat{I}_{z}^{(2)})+\gamma_cB_{x}(\hat{I}_{x}^{(1)}+\hat{I}_{x}^{(2)})\nonumber\\
 &+\frac{d_{12}}{2}\left((\hat{I}_{+}^{(1)}\hat{I}_{-}^{(2)}+\hat{I}_{-}^{(1)}\hat{I}_{+}^{(2)})-4\hat{I}_{z}^{(1)}\hat{I}_{z}^{(2)}\right),            
\end{align}
where $d_{12}=\mu_{0}\gamma_{c}^{2}/4\pi r_{12}^{3}=4$ kHz and $I_{\pm}^{(i)}=I_{x}^{(i)}\pm iI_{y}^{(i)}$. For the $m_s=+1$ manifold, the dipolar coupling ($d_{12}$) is much smaller than the hyperfine interaction ($\mathbb{A}_{z}$) and thus it is neglected. Therefore, the Hamiltonian reads
\begin{align}\label{hamilms_1_full}
 H^{m_s=+1}&=(D+\gamma_eB_z)+\gamma_cB_z(\hat{I}_{z}^{(1)}+\hat{I}_{z}^{(2)})\nonumber\\
 &\quad+\gamma_cB_{x}(\hat{I}_{x}^{(1)}+\hat{I}_{x}^{(2)})+A_{zz}^{(1)}\hat{I}_{z}^{(1)}+A_{zz}^{(2)}\hat{I}_{z}^{(2)}\nonumber\\
 &\quad+\frac{1}{2}A_{ani}^{(1)}(\hat{I}_{+}^{(1)}e^{-i\phi_{1}}+\hat{I}_{-}^{(1)}e^{i\phi_{1}})\nonumber\\
 &\quad+\frac{1}{2}A_{ani}^{(2)}(\hat{I}_{+}^{(2)}e^{-i\phi_{2}}+\hat{I}_{-}^{(2)}e^{i\phi_{2}}),
\end{align}
where $A_{ani}^{(i)}=(A_{zx}^{(i)2} +A_{zy}^{(i)2} )^{1/2}$ \cite{Jamonneau}. In order to improve readability through an easier-to-follow protocol, we only consider isotropic coupling in the hyperfine interaction for each nuclear spin ($A_{ani}^{(i)}=0$). Moreover, we choose \carb with isotropic coupling of few MHz~\cite{Nizovtsev_2014} ($A_{zz}^{(1)}=12.45$ MHz, $A_{zz}^{(2)}=2.28$ MHz), for which $A_{zz}^{(i)}\gg \gamma_cB_x,\gamma_cB_z$. We can finally write the Hamiltonian in the $m_s=+1$ manifold as
\begin{equation}\label{hamilms_1}
H^{m_s=+1}=(D+\gamma_eB_z) + A_{zz}^{(1)}\hat{I}_{z}^{(1)}+A_{zz}^{(2)}\hat{I}_{z}^{(2)}.
\end{equation}

We remark that considering the full Hamiltonian in Eq.~\eqref{hamilms_1_full} does not change our main conclusions, neither the performance of the protocol. For more details see Appendix \ref{GHamiltonian}. Furthermore, control of nuclear spins using the anisotropic hyperfine coupling have been implemented in single crystal of irradiated malonic acid \cite{Hodges08,Zhang2011}.   

In the next section we seek for a subspace in the $m_s=0$ manifold where states are well isolated from magnetic noise.

\section{Decoherence-free subspace}\label{section_decohe}

Quantum registers implemented with \carb nuclear spins in diamond are subjected to an intrinsic decoherence airisig from magnetic noise \cite{Dutt07,Maze08,Kalb2018}. The coherence time of these registers can be improve by active techniques like error-correction or decoupling pulses, but also by passive isolation. The latter, can be worked out in the framework of a Decoherence-Free Subspace (DFS)\cite{Lidar98}. DFS have been already implemented for hybrid systems given by an \NVm electron spin and one~\cite{Rama2020} or two~\cite{Reiserer2016} \carb nuclear spins. Here, we look for a logic qubit given by two states that are well separated from the rest and exhibit a flat energy distribution in terms of the external magnetic field. Firstly, we find the eigenstates of the Hamiltonian for the $m_s=0$ manifold, which are given by 
\begin{align}
 \ket{\psi_{i}}&=\frac{1}{\sqrt{2}}(\alpha_{i}+\beta_{i})\ket{\downarrow\downarrow}+\frac{1}{\sqrt{2}}(\alpha_{i}-\beta_{i})\ket{\uparrow\uparrow}\nonumber\\
 &+\frac{\xi_{i}}{\sqrt{2}}(\ket{\downarrow\uparrow}+\ket{\uparrow\downarrow}),\\
 \ket{\psi_2}&=\frac{1}{\sqrt{2}}\left(\ket{\downarrow\uparrow} - \ket{\uparrow\downarrow}\right),
\end{align}
where $i=1,3,4$ and the coefficients are
\begin{align}
 \alpha_i&=-\frac{2B_{x}\gamma_{c}(d_{12}+2E_{i})}{\sqrt{d_i}},\nonumber\\
 \beta_{i}& = \frac{4B_{x}B_{z}\gamma_{c}^{2}}{\sqrt{d_i}},\nonumber\\
 \xi_{i}&=\frac{4\gamma_{c}^{2}B_{z}^{2}-(d_{12}+2E_i)^{2}}{\sqrt{d_i}},
\end{align}
with
\begin{align}\label{d_coef}
 d_{i}&=(d_{12}+2E_{i})^{4}+4\gamma_{n}^{2}(B_{x}^{2}-2B_{z}^{2})(d_{12}+2E_{i})^{2}\nonumber\\
 &+16B_{z}^{2}\gamma_{n}^{4}(B_{x}^{2}+B_{z}^{2}).
\end{align}
The corresponding eigenvalues are: $E_{1}=2\sqrt{Q} \cos(\theta/3)$, $E_2=0$, $E_{3}= 2\sqrt{Q} \cos\left((\theta+4\pi)/3\right)$ and $E_{4}=2\sqrt{Q} \cos\left((\theta+2\pi)/3\right)$, with $\cos(\theta)=R/\sqrt{Q^3}$, $Q=(3d_{12}^{2}+4\gamma_{c}^{2}(B_{x}^{2}+B_{z}^{2}))/12$ and $R =(d_{12}^{3}+2d_{12}\gamma_{c}^{2} (B_{x}^2 - 2 B_{z}^2) )/8$. 

It is important to notice that the eigenvalue $E_2$ has zero energy, while the corresponding eigenstate ($\ket{\psi_2}$) is a maximally entangled state (Bell state). Maximally entangled eigenstates with zero eigenvalue are exceptional for quantum information, since  they are robust to random fluctuations in magnetic fields that originates dephasing noise. Preparation of such states have been already proposed for ultracold atoms \cite{Reyes2014} and cavity QED lattice \cite{Coto2016}.  Here, the ancilla spin (\NVm) that is used for control and readout, it is also well isolated from external magnetic noise as being in the $m_s=0$ state. Therefore, our logic qubit will be given by the two \carb nuclear spins, with logic states 
\begin{align}
& \ket{0_L}=\ket{\psi_2},\\
& \ket{1_L}=\ket{\psi_3}.
\end{align}

The eigenstate $\ket{\psi_3}$ is close to $\ket{\psi_2}$ while being far away from the other two eigenstates, and it is also robust against variations in the magnetic fields $B_x$ and $B_z$ as shown in Fig.~\ref{figure4}-(c) and -(d), respectively. Furthermore, by setting $\theta=\pi/2$ ($R=0$), the magnetic fields can be tuned to $B_{z}=\sqrt{d_{12}^{2}+2\gamma_c^{2}B_{x}^{2}}/2\gamma_c$ for which the corresponding eigenvalue $E_3=0$ and the two logic states become degenerated.

\begin{figure}[h]
 \centering
 \includegraphics[width=230pt]{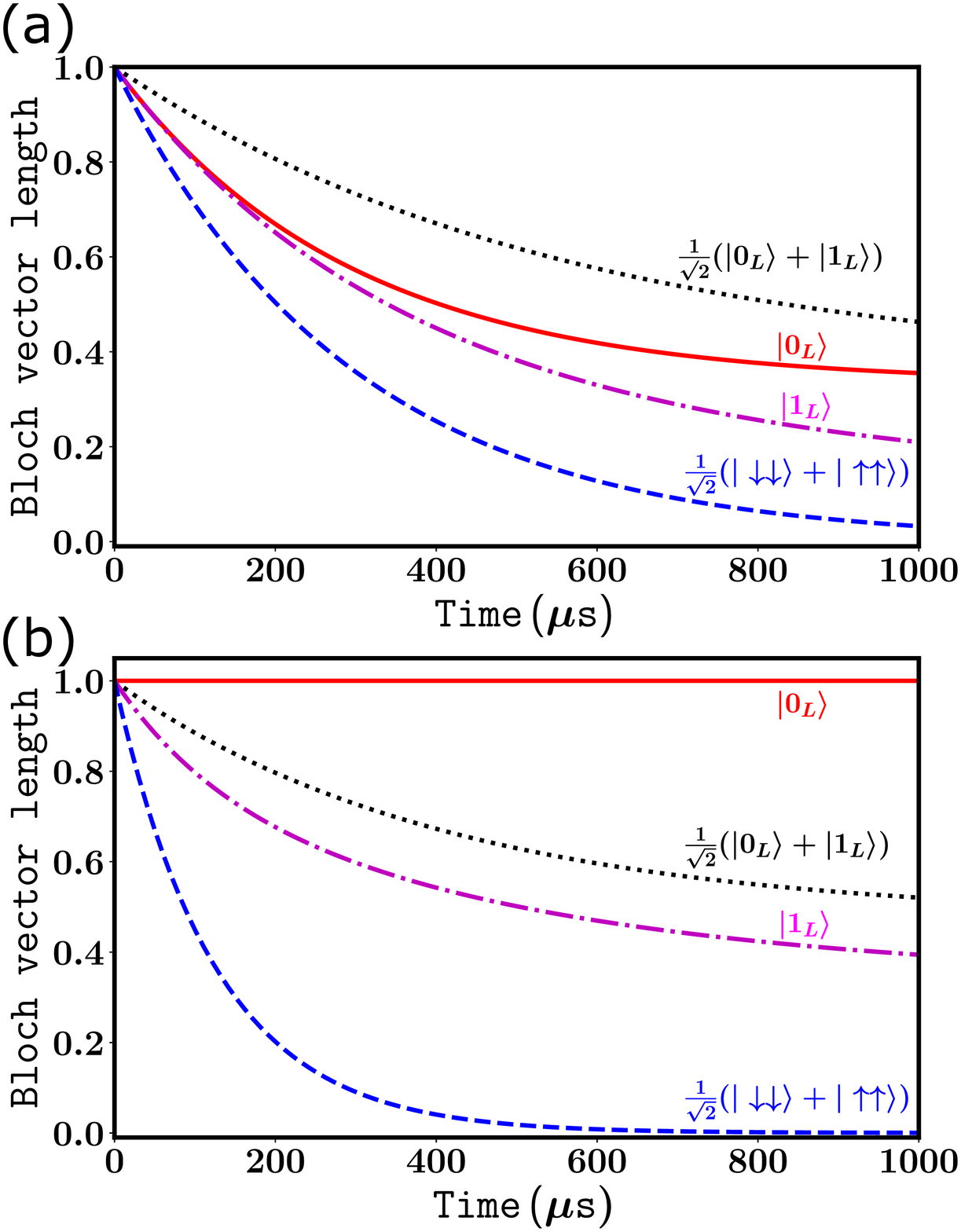}
 \caption{Bloch vector length as a function of time for a transverse relaxation process with (a) independent reservoir, (b) common reservoir. We observe that encoding information in the quantum register provides an advantage over states outside the DFS. }\label{fig_bloch}
\end{figure}

To illustrate the mitigation of decoherence due to the DFS, we consider transverse relaxation of the \NVm and the two \carb. The characteristic time $T_2^\ast$ for the \NVm is around few microseconds \cite{Childress281,Blok}, while for the \carb is hundreds of microseconds \cite{Dutt07}. Hereafter, we set $T_2^\ast=7$ $\mu$s, $T_{2n_{1}}^\ast=500$ $\mu$s and $T_{2n_{2}}^*=700$ $\mu$s. This relaxation process is modeled through a markovian master equation given by
 \begin{align}\label{ME_in}
 \frac{d\rho(t)}{d t} &= -i[ H^{m_s},\rho]+(1/T_2^*)\left(2S_z\rho S_z-S_z^2\rho-\rho S_z^2\right) \\ \nonumber &+ \mathcal{L}_I, 
\end{align}
%considering two scenarios, namely: (\textit{i}) ~\eqref{ME_in}, (\textit{ii}) ~\eqref{ME_co}, 

where $\mathcal{L}_I=\sum_{i=1}^2(1/T_{2n_{i}}^*)\left(2 I_z^{(i)}\rho I_z^{(i)}-I_z^{(i)^2}\rho-\rho I_z^{(i)^2}\right)$ considers independent reservoirs, and $\mathcal{L}_I= (1/T_{2n}^*)\left(2 \mathbb{I}_z\rho \mathbb{I}_z-\mathbb{I}_z^2\rho-\rho \mathbb{I}_z^2\right)$ considers a common reservoir, with $\mathbb{I}_z=I_z^{(1)}+I_z^{(2)}$ and $T_{2n}^*=500$ $\mu$s. In Fig.~\ref{fig_bloch} we show the Bloch vector length $(\mean{\sigma_x}^2+\mean{\sigma_y}^2+\mean{\sigma_z}^2)^{1/2}$ as a function of time for different preparation of the logic qubit $\{\ket{0_L},\ket{1_L}\}$. For comparison, we consider another logic qubit $\{\ket{\Downarrow}=\ket{\downarrow\downarrow},\ket{\Uparrow}=\ket{\uparrow\uparrow}\}$ that is out of the DFS. We remark that the qubit encoded in the DFS leads to decreased dephasing, therefore, yielding enhancement in coherence time. Furthermore, we observe that a common reservoir improves dephasing for the qubit encoded in DFS, while it increases (deteriorates) dephasing for the other qubit. 

We have proved that encoding information in DFS through a quantum register (logical qubit $\{\ket{0_L},\ket{1_L}\}$) provides an advantage over a qubit outside the DFS.  The remaining challenge is to prepare and control it. In what follows, we will focus on preparing the system in the DFS and performing spin-flip operation on the logic qubit.

\subsection{Accessing the Decoherence-free subspace}\label{accessing_DFS}

To begin with, we remark that the Hamiltonian for the $m_s=+1$ manifold in Eq~\eqref{hamilms_1} is diagonal in the bare basis with eigenstates: $\ket{\psi_5}=\ket{\downarrow\downarrow}$, $\ket{\psi_6}=\ket{\downarrow\uparrow}$, $\ket{\psi_7}=\ket{\uparrow\downarrow}$, $ \ket{\psi_8}=\ket{\uparrow\uparrow}$. Moreover, we consider two driving fields upon \NVm given by $\sqrt{2}\Omega_p\cos(\omega_pt)S_x$ and $\sqrt{2}\Omega_s\cos(\omega_st)S_x$, with $\Omega_p(t)=\Omega_{0}\exp\left(- (t-t_d/2)^2/2\sigma^2\right)$ and $\Omega_s(t)=\Omega_{0}\exp\left(- (t+t_d/2)^2/2\sigma^2\right)$. $\Omega_0$ is the Rabi frequency, $t_d$ is the delay between pulses, and $\sigma$ is the width of the Gaussian. The total Hamiltonian ($\lbrace m_s=0,m_s=+1\rbrace$) in a multi-rotating frame reads
\begin{align}
 \hat{H}&= \delta \hat{\sigma}_{22} + E_3\hat{\sigma}_{33}+E_4\hat{\sigma}_{44}+\Delta_3\hat{\sigma}_{55} \nonumber \\ &+ \Delta_1\hat{\sigma_{66}} +\Delta_4\hat{\sigma}_{77}+\Delta_5\hat{\sigma}_{88}\nonumber\\
 &+ \frac{\Omega_{p}(t)}{2}\left(\chi_{51}\hat{\sigma}_{51}+\chi_{61}\hat{\sigma}_{61}+\chi_{71}\hat{\sigma}_{71}+\chi_{81}\hat{\sigma}_{81}\right)\nonumber\\
 &+\frac{\Omega_{s}(t)}{2}\left(\chi_{62}\hat{\sigma}_{62}+\chi_{72}\hat{\sigma}_{72}\right),
\end{align}
where $\delta = \Delta_1-\Delta_{2}$ corresponds to the two-photon detuning and $\Delta_1 = E_6-E_1-\omega_{p}$, $\Delta_2= E_6-E_2-\omega_{s}$,  $\Delta_3 = E_5-E_1-\omega_{p}$, $\Delta_4 = E_7-E_1-\omega_{p}$, $\Delta_5 = E_8-E_1-\omega_{p}$ are one-photon detuning. $\omega_{p(s)}$ are the frequencies of the driving fields, $\sigma_{ij}=\vert\psi_i\rangle\langle \psi_j\vert $ with $\vert \psi_i\rangle$ the eigenstates ($i=1\dots 8$), and the coefficients are $\chi_{ij}=\langle \psi_i\vert\hat{V}\vert\psi_{j}\rangle$, with $\hat{V}=\vert1\rangle\langle 0\vert + \vert1\rangle\langle 0\vert$. 

\begin{figure*}[ht]
\centering
\includegraphics[scale=0.18]{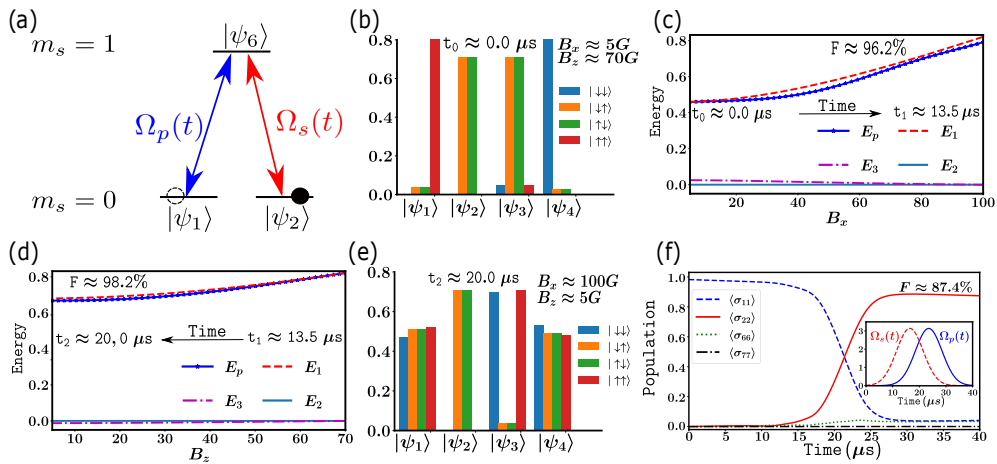}
\caption{Preparation of the system in the decoherence-free subspace. (a) Energy levels of relevant states. (b) Probability amplitude of each eigenstate at $t =0$. The state $\ket{\psi_1} $ is projected to $ \ket{0\uparrow \uparrow} $. (c) Adiabatic evolution of the state $\ket{\psi_{1}} $, as shown by the mean energy $E_p=\mbox{Tr}[\rho(t) H^{m_s=0}]$, varying $ B_{x}(t) = \epsilon_{x} t + B_{x0} $, with $ \epsilon_{x} = 7 $ G/$\mu$s and $ B_{x0} = 5$ G, while the $B_z$ remains constant at $ B_{z} = 70$ G. (d) Subsequent adiabatic evolution of the eigenstate $ \ket{\psi_{1}} $ after evolution in (c), varying $ B_{z}(t) = \epsilon_{z} t + B_{z0} $, with $ \epsilon_{z} = - 10 $ G/$\mu s$ and $ B_{z0} = 70$ G, while the component $ B_{x} = 100$ G remains constant. The overall time is $t = 20.0~\mu$s \cite{Note1}. (e) Probability amplitude for each eigenstate after the second adiabatic evolution in (d), note that $\ket{\psi_1}$ has a balanced contribution of all bare states. With this configuration we reach the highest fidelity to prepare the $\ket {\psi_{2}}$. (f) Population transfer via STIRAP of the state $\ket{\psi_1}$ to  $\ket{\psi_2}$ with fidelity $87.4\%$.  $\sigma=5$ $\mu s$, $\Omega_{0}/2\pi=0.5$ MHz. The Inset shows the pulse sequence; Stokes pulse $(\Omega_{s})$ precedes the pump pulse $(\Omega_{p})$.}\label{figure4}
\end{figure*}

Hereafter, we consider the transverse relaxation process for independent reservoirs given in Eq.~\eqref{ME_in}. Our first goal is to prepare the system in a maximally entangled eigenstate $\ket{\psi_2}$ via microwave pulses, harnessing the hyperfine coupling between the \NVm and \carb nuclear spins. In Fig.~\ref{figure4} (a) we show the energy levels of relevant states (initial $\ket{\psi_1}$, target $\ket{\psi_2}$ and intermediate $\ket{\psi_6}$) used in our protocol. Firstly, we initialize the system in the state $\ket{0\uparrow\uparrow}$. Polarization of the \NVm and multiples nuclear spins have been successfully addressed~\cite{Jacques09,Taminiau2014,Bradley}. We find that the eigenstate $\ket{\psi_1}$ is closely projected to $\ket{0\uparrow\uparrow}$ by appropriately choosing the external magnetic fields, see Fig.~\ref{figure4}~(b). The population transfer to state $\ket{\psi_2}$ is performed via the intermediate state $\ket{\psi_6}$ ($\ket{\psi_1}\rightarrow\ket{\psi_6}\rightarrow\ket{\psi_2}$). At this stage, different protocols for population transfer fail, due to the weak perpendicular component of the magnetic field $B_x$ that prevents nuclear spin-flip. Therefore, we adiabatically increase the magnetic field up to $B_x=100$~G, see Fig.~\ref{figure4}~(c), leaving $B_z$ constant. Next, we consider an intuitive pulse sequence, as detailed in Appendix~\ref{pul_int}, that transfers population to state $\ket{\psi_6}$ and then to state $\ket{\psi_2}$. This process, which takes around $30$~$\mu$s, only reaches a small fidelity ($F=\mbox{Tr}[\rho\vert\psi_2\rangle\langle\psi_2\vert]$) of $52\%$. This fidelity is strongly limited by the dephasing rate of the \NVm electron spin, and it deteriorates as the population in the excited state ($\ket{\psi_6}$) increases. In order to overcome this drawback we consider the \textit{Stimulated Raman Adiabatic Passage} (STIRAP)~\cite{Gaubatz90,Bergmann,Bergmann15,Vitanov}, which adiabatically transfers population from state $\ket{\psi_1}$ to $\ket{\psi_2}$ without increasing the population in state $\ket{\psi_6}$. In this way, we decrease decoherence and gain in selectivity, avoiding cross talk with other excited states. Nevertheless, in this configuration of magnetic fields STIRAP underperforms because of the small component of $\ket{0\downarrow\uparrow}$ in state $\ket{\psi_1}$ (Appendix~\ref{pul_int}). Henceforth, we apply a second adiabatic evolution (Fig.~\ref{figure4}~(d)) where the magnetic field $B_z$ is decreased down to $5$~G while $B_x$ remains constant ($100$~G). This evolution drives state $\ket{\psi_1}$ towards a more balanced contribution of all bare states (Fig.~\ref{figure4}~(e)), enabling STIRAP. We remind that the components of the target state $\ket{\psi_2}$ do not depends on the external magnetic field. We now apply the STIRAP sequence, as illustrate in Fig.~\ref{figure4}~(f). We observe that population transfer improves, reaching a fidelity of $F=87.4\%$. To account for the total time, we consider the overall adiabatic evolution time ($\approx 20$ $\mu$s) plus the STIRAP time ($\approx 30$ $\mu$s), that yields a total time of $50$ $\mu$s. Note that we do not consider here dead time for initialization.

To conclude, we are able to prepare a Bell's state ($\ket{\psi_2}$), which is one of the states of our decoherence-free quantum register, in an all-microwave setup by manipulating nominal non-nuclear-spin-preserving transitions ($\ket{\psi_1}\rightarrow\ket{\psi_6}$). This is possible due to the dipolar interaction and the magnetic field $B_x$ that set a different quantization axis in $m_s=0$ with respect to $m_s=+1$ manifold.  In this way, we avoid the much slower process of using radiofrequency (rf) to control nuclear spins transitions, which is limited by the small gyromagnetic ratio $\gamma_{c}/2\pi = 1.07 $ kHz/G, that is three order of magnitude smaller than that for the electron spin. For instance, preparation of Bell-like states have been recently achieved combining phase-controlled rf interleaved with electron spin dynamical decoupling with gate time ranging from hundreds to thousands of microseconds~\cite{Bradley}, in contrast to the expected fifty microseconds reported here. 

%Our preparation time can be further improved by circumventing the adiabatic condition and counteracting the effect of the loss of adiabaticity with an external control. To achieve this goal we use sequence proposed in \cite{PhysRevLett.116.230503}, named MOD-SATD. \rc{ESTO QUE VIENE HABRIA QUE ELIMINARLO DADO QUE MOD-SATD NO ESTA DANDO MUCHO BENEFICIO. }In this part we have that when the adiabatic condition is fulfilled, this protocol loses robustness, although for short times the MOD-SATD is more efficient than the STIRAP as can be seen in the Fig ~\ref{Figure6}.
%
%\begin{figure}[h]
%\centering
%\includegraphics[width=230pt]{Figure6}
%\caption{ MOD-SATD protocol vs STIRAP, considering dissipation only in the electronic spin by means of a pure dephasing. $\Gamma =1/T_{2}^{\ast}$, $T_{2}^{\ast}\approx7\mu s$. MOD-SATD is more efficient at early times, while STIRAP is more robust when the adibatic process is slower.}\label{Figure6}
%\end{figure}

\subsection{Performing spin-flip operation on logic qubit}

Manipulating states that belong to a DFS is  a difficult task, that is hindered by the control sequence itself. The available control pulses at hand or even imperfect fidelity of the control sequence may take the encoded information out of the DFS. Towards logic operations that hold the information in the DFS, a universal set of logical manipulations based on decoupling pulses and exploiting symmetries have been proposed \cite{Fortunato2002,Cappellaro2006}. Nevertheless, in this work we follow a different approach that harnesses the selectivity of the STIRAP sequence to avoid unwanted cross talk to other states. We consider a $\Lambda$-configuration $\lbrace \ket{0_L},\ket{1_L},\ket{\psi_6} \rbrace$ and by applying resonant pulses $\Omega_p(t)$ and $\Omega_s(t)$ for driving transitions $\ket{0_L} \leftrightarrow \ket{\psi_6}$ and $\ket{1_L}\leftrightarrow \ket{\psi_6}$, respectively, we are able to perform a spin-flip operation on the logic qubit. In order to maximize the fidelity, we first adiabatically increase the magnetic field $B_z$ up to $70$~G, which makes the logic states nearly degenerated. 

\begin{figure}[h]
 \centering
 \includegraphics[width=230pt]{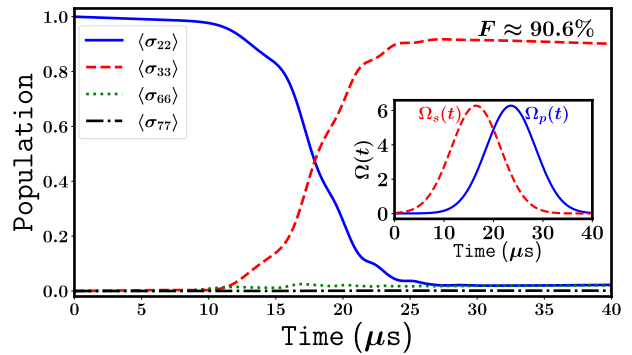}
 \caption{Population transfer via STIRAP of the state $\ket{\psi_2}$ to $\ket{\psi_{3}}$ with $90.6\%$ fidelity.  $\sigma=5$ $\mu s$, $\Omega_{0}/2\pi=1$ MHz, $B_x=100$~G and $B_{z}\approx70$~G. The panel above shows the STIRAP pulse sequence; Stokes pulse $(\Omega_{s})$ precedes the pump pulse $(\Omega_{p})$.}\label{DFS}
\end{figure}

In Fig.~\ref{DFS} we show the population transfer (spin-flip operation) from $\ket{0_L}$ to $\ket{1_L}$ achieving a fidelity of $ 90.6\% $. This high fidelity results from the negligible population in states that belongs to the $m_s=+1$ manifold, e.g. $\ket{\psi_6}$ and $\ket{\psi_7}$. We remark that we are not using dynamical decoupling or pulse shaping, which could improve further our gate. The operation in the opposite direction ($\ket{1_L} \rightarrow\ket{0_L}$) can be also attained via STIRAP by interchanging the pulse order, yielding the same fidelity. Moreover, one can distinguish between opposite directions upon the logic qubit gate by leaving the pulse order fixed. On the one hand, when the first pulse acts on the state with no occupation, which creates the dark state that STIRAP relies on (see Fig.~\ref{DFS}), no excitation reaches the $m_s=+1$ manifold. On the other hand, when the first pulse triggers the transition with full occupation, which is known as b-STIRAP \cite{Klein07}, a small excitation reaches $m_s=+1$. By applying a green laser during each sequence, fluorescence will allows to distinguish the STIRAP path (higher fluorescence) from the b-STIRAP path (lower fluorescence). We note that the green laser induces depolarization in the $m_s=0$ manifold \cite{Jamonneau,Coto2017}, which must be considered for applications.

\section{Controlling a single \carb register}\label{section_single}

\begin{figure}[ht]
\centering
\includegraphics[scale=0.22]{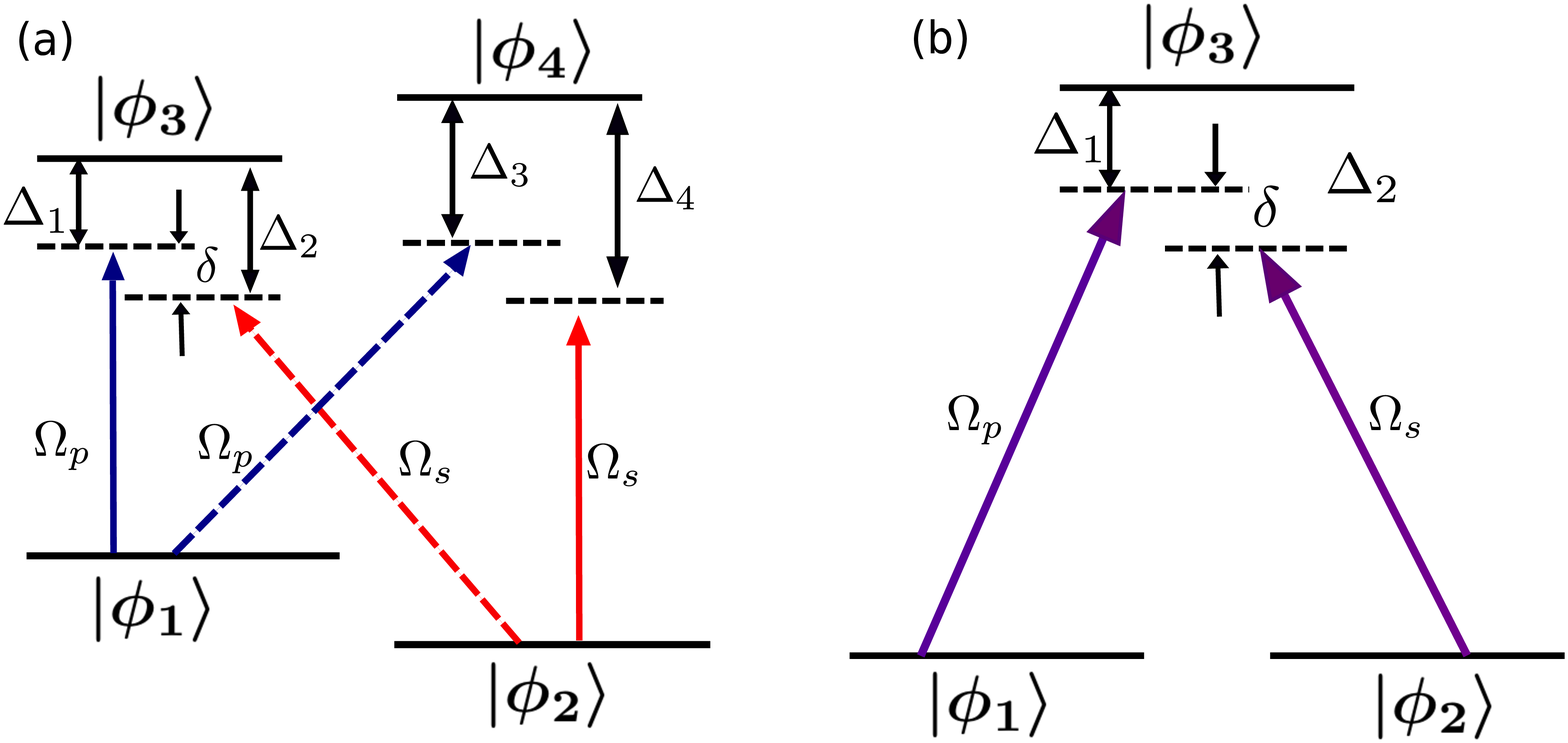}
\caption{(a) Bi-partite system given by an \NVm electronic spin hyperfine coupled to a nearby \carb nuclear spin. It shows all the possible transitions through the pump (blue) and stoke (red) fields and they are adjusted in such a way that the state $\ket{\phi_{4}} $ is not reached to excite and reduce it to a system of three levels.  (b) Effective $\Lambda$-system driven by a pump and Stokes microwave pulses.}
\label{fig2}
\end{figure}

In this section we focus on controlling a single \carb nuclear spin that is hyperfine coupled to an \NVm. In a previous work~\cite{Coto2017}, the authors proposed the STIRAP sequence to coherently control the \carb. However, such control relied on the anisotropy of the hyperfine coupling, which varies among different samples. Towards a universal protocol, we replace here the anisotropy ($A_{ani}=0$) by a magnetic field ($B_x$) that is perpendicular to the \NVm quantization axis, such that when the electron spin is in the state $m_s=0$ the quantization axis for the nuclear spin will follow the magnetic field. This means that the eigenstates in the submanifold $m_s=0$ will be mixed (see Eqs.~\eqref{eigen1}-\eqref{eigen2}), which enables non-nuclear-spin-preserving transitions (non-NSPT). When the electron spin is in state $m_s=+1$, we neglect $\gamma_cB_x$ since it is much smaller than the hyperfine coupling ($A_{zz}\approx1.07$~MHz). Therefore, the quantization axis for the nuclear spin will be given by ${A}_{zz}$, which means that the eigenstates in the $m_s=+1$ submanifold are not mixed (see Eqs.~\eqref{eigen3}-\eqref{eigen4}). For this bi-partite system, the Hamiltonian reduces to
\begin{eqnarray}\label{hamilt_bi}
H&=&DS_{z}^{2}+\gamma_{e}B_zS_z+\gamma_{n}B_zI_z+\frac{1}{2}\gamma_{n}B_{x}(I_{+}+I_{-})\nonumber\\
 &+&A_{zz}S_zI_z.
 \end{eqnarray}
For convenience, we restrict our analysis to the $\{m_s=0,m_s=+1\}$ manifold. The eigenstates of the above Hamiltonian are,
\begin{align}
\label{eigen1}
 |\phi_1\rangle&=\cos(\theta)|0\uparrow\rangle+\sin(\theta)\ket{0\downarrow},\\
\label{eigen2}
|\phi_2\rangle&=\sin(\theta)|0\uparrow\rangle-\cos(\theta)\ket{0\downarrow},\\
 \label{eigen3}
  |\phi_3\rangle&=|1\uparrow\rangle,\\
  \label{eigen4}
 |\phi_4\rangle&=|1\downarrow\rangle.
\end{align}
The mixing angle is defined through $\tan(\theta)=B_x/(B+B_z)$, with $B=\sqrt{B_x^2+ B_z^2}$. We remark that $\theta$ allows us to control the strength of non-NSPT. For instance, in the absence of the perpendicular magnetic field ($\theta=0$), non-NSPT are forbidden under microwave (MW) pulses, see dashed lines in Fig.~\ref{fig2} (a). However, with a balanced mixing of the states in $m_s=0$, $\theta\approx0.74$~rad with $B_x=100$~G and $B_z=10$~G, we can set the $\Lambda$-configuration shown in Fig.~\ref{fig2} (b). This $\Lambda$-system enables fast coherent control over the nuclear spin \cite{Coto2017,Huillery}, via selective electron spin transitions that are driven by MW pulses.

 We now write the total Hamiltonian in the eigenstate basis $\{|\phi_1\rangle,|\phi_2\rangle,|\phi_3\rangle,|\phi_4\rangle\}$ and in a multi-rotating frame, see Appendix~\ref{rot_frame}, such that
\begin{align}
 \tilde{H}&=\delta\sigma_{22}+\Delta_{3}\sigma_{33}+\Delta_{4}\sigma_{44}\nonumber\\
 &+\frac{\Omega_{p}(t)}{2}\left(\cos(\theta)\sigma_{31}+\sin(\theta)\sigma_{41}+h.c\right)\nonumber\\
 &+ \frac{\Omega_{s}(t)}{2}\left(\sin(\theta)\sigma_{32}-\cos(\theta)\sigma_{42}+h.c\right)
\end{align}
where $\Delta_2=(E_3-E_2-\omega_s)$, $\Delta_3=E_3-E_1-\omega_p$ and $\Delta_4=E_4-E_1-\omega_p$ are the one-photon detunings, and $\delta=\Delta_3-\Delta_2$ is the two-photon detuning, as shown in Fig.~\ref{fig2} (a). For completeness, we include in our calculations the state $\ket{\phi_4}$. However, we shall see that the population in this state is negligible during the pulse sequence, which support our choice of $\Lambda$-system in Fig.~\ref{fig2} (b). We set resonant driven fields $\omega_p=E_3-E_1$ and $\omega_s=E_3-E_2$, that yields  $\Delta_2=\Delta_3=\delta=0$.

\begin{figure}[ht]
\centering
\includegraphics[width=240pt]{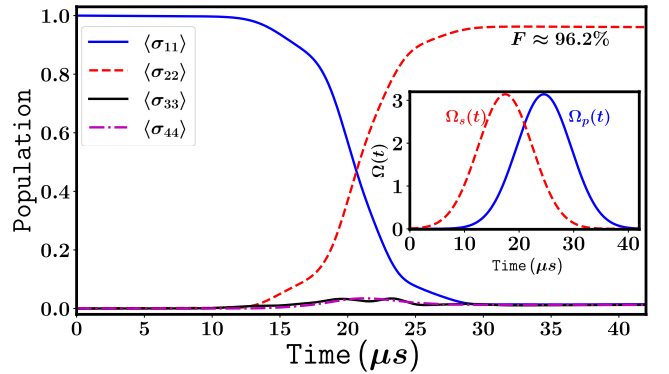}
\caption{Population transfer from the state $\ket{\phi_1}$ to  $\ket{\phi_2}$ with $96\%$ fidelity. The parameters are $\sigma=9$ $\mu$s, $\Omega_{0}/2\pi=0.5$ MHz, $B_x=100$ G, $B_z=10$ G, $T_{2}^{\ast}=7\mu$s and $T_{2n}^{\ast}=500\mu$s . }
\label{fig3}
\end{figure}

The population transfer is strongly limited by selectivity and decoherence. The former, imposes a constraint on the Rabi frequency due to power broadening, which limits the control speed. We choose MW pulses as Gaussian fields, $\Omega_p(t)=\Omega_{13}\exp\left(- (t-t_d/2)^2/2\sigma^2\right)$ and $\Omega_s(t)=\Omega_{23}\exp\left(- (t+t_d/2)^2/2\sigma^2\right)$, and numerically found that for $\Omega_{13}=\Omega_{23}=\Omega_{0}$ and $\Omega_{0}/2\pi=0.5 $ MHz all transitions can be selectively addressed. The latter, requires control time bellow the coherence time, and the transverse relaxation process is modeled by the master equation in Eq.~\eqref{ME_in} for a single \carb. To overcome these limitations, STIRAP improves selective and fast coherent population transfer by properly setting the MW fields. Moreover, quantum interference prevents population in states that experiences high decoherence. In what follows, we focus on transferring population from an initially polarized state $\ket{\phi(0)}=\ket{\phi_1}$ to $\ket{\phi_2}$ via STIRAP. To achieve maximum fidelity the time delay is set to $t_d = \sqrt{2}\sigma$. In Fig.~\ref{fig3} we show the time evolution of the populations ($\mean{\sigma_{ii}}=\mbox{Tr}[\rho(t)\vert\phi_i\rangle\langle\phi_i\vert]$) in all states when applying the STIRAP sequence. One can observe that our protocol is suitable for transferring population in an all-microwave setup, reaching a fidelity of $96\%$ in $30$~$\mu$s. Furthermore, the off-axis magnetic field allows us to set a universal protocol that does not rely on the anisotropy of the hyperfine coupling, generalizing the results in Ref. \onlinecite{Coto2017}. Our protocol takes advantage of the faster electron spin transition to polarize the nuclear spin via non-nuclear-spin-preserving transitions. The expected control time is up to two orders of magnitude shorter than those using rf fields to control individual nuclear spins \cite{Jiang09,Bradley}.

\section{Conclusions}\label{section_conclu}

In summary, we have proposed a quantum register that lies in a decoherence-free subspace to be implemented in color centers in diamond. By fixing the \NVm electron spin in the state $m_s=0$, we prepared the two logic states of the register with two nearby \carb nuclear spins. We remark that one of these logic states is the singlet state. Furthermore, by tuning an off-axis magnetic field we enabled non-nuclear-spin-preserving transitions that we used for preparing the singlet state through Stimulating Raman Adiabatic Passage, yielding an improvement in time up to two orders of magnitud smaller than protocols based on rf fields. We also used this sequence to manipulate the quantum register and an individual nuclear spin.   

\section{acknowledgments} We thank J.~R. Maze for encouraging discussions. FJG acknowledges support from Universidad Mayor through the Doctoral fellowship. RC acknowledges support from Fondecyt Iniciaci\'on No. 11180143.

\appendix

\section{General Hamiltonian for $m_{s}=+1$}\label{GHamiltonian}

In the main text we simplified the Hamiltonian in the $m_s = +1$ manifold, yielding a simpler Hamiltonian in Eq.\eqref{hamilms_1}. Nevertheless, in this section we study the general case in Eq.\eqref{hamilms_1_full},
\begin{align}
 \hat{H}_{m_s=1}&=\omega_{e}+\omega_{1z}\hat{I}_{z}^{(1)}+\omega_{2z}\hat{I}_{z}^{(2)}+\omega_{1x}\hat{I}_{+}^{(1)}+\omega_{1x}^{\ast}\hat{I}_{-}^{(1)}\nonumber\\
 &\quad+\omega_{2x}\hat{I}_{+}^{(2)}+\omega_{2x}^{\ast}\hat{I}_{-}^{(2)},
\end{align}

where we explicitly write $\hat{I}_{x}=(\hat{I}_{+}+\hat{I}_{-})/2$ and
\begin{align}
 \omega_{e}&=D+\gamma_eB_z,\nonumber\\
 \omega_{1z}&=\gamma_nB_z+A_{zz}^{(1)},\nonumber\\
 \omega_{2z}&=\gamma_nB_z+A_{zz}^{(2)},\nonumber\\
 \omega_{1x}&=\frac{1}{2}\left(\gamma_nB_x+A_{ani}^{(1)}e^{-i\phi_{1}}\right),\nonumber\\
 \omega_{2x}&=\frac{1}{2}\left(\gamma_nB_x+A_{ani}^{(2)}e^{-i\phi_{2}}\right).
\end{align}

For simplicity, we set $\phi_1=\phi_2=0$, leading to $\omega_{1x}^{\ast}=\omega_{1x}$ and $\omega_{2x}=\omega_{2x}^{\ast}$. We now use the bare basis $\{\ket{\downarrow\downarrow},\ket{\downarrow\uparrow},\ket{\uparrow\downarrow},\ket{\uparrow\uparrow}\}$
 to write the Hamiltonian in a matrix form,
 \begin{widetext}
\begin{align}
\hat{H}^{m_s=+1}=& \begin{pmatrix}
  \omega_{e}-\frac{1}{2}(\omega_{1z}+\omega_{2z}) & \omega_{2x}       & \omega_{1x}      & 0  \\
 \omega_{2x} &   \omega_{e}-\frac{1}{2}(\omega_{1z}-\omega_{2z})        &  0    &  \omega_{1x}  \\
 \omega_{1x} & 0        &\omega_{e}+\frac{1}{2}(\omega_{1z}-\omega_{2z})    & \omega_{2x}  \\
 0 & \omega_{1x}      & \omega_{2x}    &  \omega_{e}+\frac{1}{2}(\omega_{1z}+\omega_{2z})       \\
   \end{pmatrix}.\label{hamil_mso}
\end{align}
\end{widetext}

The parameters for the hyperfine couplings are taken from~\cite{Nizovtsev_2014}, and shown in Table~\ref{Table1}. Other parameters are $\gamma_e/2\pi=2.8$ MHz, $\gamma_n/2\pi=1.07$ kHz, $D/2\pi=2.87$ GHz, $B_z=5$ G y $B_x=100$ G, we carry out STIRAP sequence in Fig.~\ref{figure4}~(f) ($\ket{\psi_1}\longrightarrow\ket{\psi_6}\longrightarrow\ket{\psi_2}$) and obtain the same fidelity of $87.4\%$.

\begin{table}
\caption{\label{Table1} Hyperfine couplings and characteristic time ($T_2^\ast$) used in our simulations. }
\begin{ruledtabular}
\begin{tabular}{cccc}
 nuclei & $A_{zz}$ (MHz) & $A_{ani}$ (MHz) & $T_{2n}^\ast$ ($\mu$s) \\
\hline
$^{13}C_1$ & $12.45$   & $1.16$ & $500$ \\
$^{13}C_2$& $2.28$   & $0.24$ & $700$ \\
\end{tabular}
\end{ruledtabular}
\end{table}

\section{Intuitive pulse sequence to prepare the entangled state}\label{pul_int}

\begin{figure}
\centering
\includegraphics[scale=0.20]{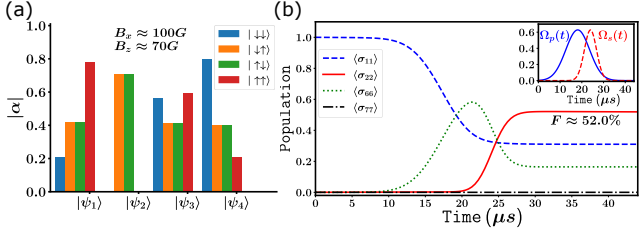}
\caption{(a) Coefficients for the eigenstates with $ B_ {x} = 100 $ G and $ B_ {z} = 70 $ G. (b) Population transfer using an intuitive pulse sequence (inset) with  $52\%$ success. $\sigma_{p} = 5.5 \mu$s, $\sigma_{s} = 2.8 \mu$s, $\Omega_0/2\pi=0.1$ MHz ensures no population in other excited states.} \label{Fig7}
\end{figure}

In this Section we provide more details about the intuitive pulse sequence mentioned in Section~\ref{accessing_DFS} to prepare the state $\ket{\psi_2}$. First, we consider the $\theta$ introduced below Eq.~\eqref{d_coef} through $\cos(\theta)=R/\sqrt{Q^3}$, with $Q=(3d_{12}^{2}+4\gamma_{c}^{2}(B_{x}^{2}+B_{z}^{2}))/12$ and $R =(d_{12}^{3}+2d_{12}\gamma_{c}^{2} (B_{x}^2 - 2 B_{z}^2) )/8$. By setting $\theta=\pi/2$, we obtain the following relation between the magnetic fields, 
\begin{equation}\label{Bz}
B_{z}=\frac{1}{2\gamma_c}\sqrt{d_{12}^{2}+2\gamma_c^{2}B_{x}^{2}}.
\end{equation}
We now replace $B_x = 100$~G and $d_{12}=4$~kHz, that yields $B_z \approx 70$~G. This configuration of magnetic fields results in state $\ket{\psi_3}$ having zero eigenvalue. The probability amplitude for the components of the eigenstates are depicted in Fig.~\ref{Fig7} (a). The intuitive sequence consist in two Gaussian pulses where width and delay have been optimized to increase the fidelity of the population transfer from state $\ket{\psi_{ 1}}$ to $ \ket{\psi_{2}}$. The process takes around $30 \mu$s with a small fidelity of $52 \%$, which is strongly affected by the transversal relaxation of the electron spin, see Fig. \ref{Fig7} (b). For this reason we use STIRAP instead, which provides better fidelity. %For this reason, it is necessary to make another adiabatic evolution such that now the component of the magnetic field in $ x $ remains constant and the component $ z $ of the field is decreased, obtaining a greater fidelity of preparation but in a time that is approximately the double that obtained with this protocol. Even so, this is still a very efficient model compared to other models where they use RF that takes around 600 $ \mu s $ to control the nuclear spin bath \cite{Jiang09,Brown11}.

\section{Single \carb Hamiltonian and Multi-Rotating Frame}\label{rot_frame}

The total Hamiltonian can be written as $H_t=H + H_i$, where
\begin{align}\label{H_apen}
H&=DS_{z}^{2}+\gamma_{e}B_zS_z+\gamma_{n}B_zI_z+\frac{1}{2}\gamma_{n}B_{x}(I_{+}+I_{-})\nonumber\\
 &+A_{zz}S_zI_z,
 \end{align}
 and 
 \begin{equation}\label{Hi_apen}
 H_i = \sqrt{2}\Omega_p\cos(\omega_pt)S_x + \sqrt{2}\Omega_s\cos(\omega_st)S_x,
 \end{equation}
 with $\Omega_i$ and $\omega_i$ the Rabi frequency and MW field frequency, respectively. $S_x$ is the electron spin-$1$ operator. The eigenvalues and eigenstates of the Hamiltonian $H$~(\ref{H_apen}) are given by, 
 \begin{align}
   E_1 &=\frac{1}{2} \gamma_c B,\\
   E_2 &=-\frac{1}{2}\gamma_c B,\\
   E_3 &=D+\gamma_eB_z+\frac{1}{2}\sqrt{(\gamma_{n}B_x)^{2}+(A_{zz}+\gamma_cB_z)^{2}},\\
   E_4 &=D+\gamma_eB_z-\frac{1}{2}\sqrt{(\gamma_{n}B_x)^{2}+(A_{zz}+\gamma_cB_z)^{2}},\\
   E_5 &=D-\gamma_eB_z+\frac{1}{2}\sqrt{(\gamma_{n}B_x)^{2}+(A_{zz}+\gamma_cB_z)^{2}},\\
   E_6 &=D-\gamma_eB_z-\frac{1}{2}\sqrt{(\gamma_{n}B_x)^{2}+(A_{zz}+\gamma_cB_z)^{2}},
  \end{align}

\begin{align}
 |\phi_1\rangle&=\cos(\theta)|0\uparrow\rangle+\sin(\theta)\ket{0\downarrow},\\
 |\phi_2\rangle&=\sin(\theta)|0\uparrow\rangle-\cos(\theta)\ket{0\downarrow},\\
 |\phi_3\rangle&=|+1\uparrow\rangle,\\
 |\phi_4\rangle&=|+1\downarrow\rangle,\\
 |\phi_5\rangle&=|-1\uparrow\rangle,\\
 |\phi_6\rangle&=|-1\downarrow\rangle,
 \end{align}
  with $B=\sqrt{B_z^2+B_x^2}$ and  $\tan(\theta)=B_x/(B+B_z)$. For the eigenstates, we have neglected the contribution of $\gamma_cB_x$ in the subspace $\lbrace m=\pm1\rbrace$, given that $A_{zz}\gg\gamma_cB_x$. We now transform the total Hamiltonian in the eigenstate basis as   
 \begin{widetext}
 \begin{equation}
 H_t = \begin{pmatrix}
 E_1 & 0 & \cos(\theta)\Omega(t) & \sin(\theta)\Omega(t)  & \cos(\theta)\Omega(t)  & \sin(\theta)\Omega(t)  \\
 0 & E_2 & \sin(\theta)\Omega(t) & -\cos(\theta)\Omega(t)  & \sin(\theta)\Omega(t) & -\cos(\theta)\Omega(t) \\
 \cos(\theta)\Omega(t) & \sin(\theta)\Omega(t) & E_3 &0 &0 &0 \\
 \sin(\theta)\Omega(t) & -\cos(\theta)\Omega(t) &0 & E_4 & 0 &0 \\
 \cos(\theta)\Omega(t) & \sin(\theta)\Omega(t) &0 &0 &E_5 &0\\
 \sin(\theta)\Omega(t) & -\cos(\theta)\Omega(t) &0 &0 &0 &E_6 \end{pmatrix},
 \end{equation}
 \end{widetext}
 where $\Omega(t)=\left(\Omega_p\cos(\omega_pt)+\Omega_s\cos(\omega_st)\right)$. The desired Hamiltonian in the $\lbrace m_s=+1,m_s=0 \rbrace$ manifold is obtained by truncating this matrix,
 \begin{equation}
 H_t = \begin{pmatrix}
 E_1 & 0 & \cos(\theta)\Omega(t)  & \sin(\theta)\Omega(t)  \\
 0 & E_2  & \sin(\theta)\Omega(t) & -\cos(\theta)\Omega(t) \\
  \cos(\theta)\Omega(t) & \sin(\theta)\Omega(t) &E_3 &0\\
 \sin(\theta)\Omega(t) & -\cos(\theta)\Omega(t) &0 &E_4 \end{pmatrix}.
 \end{equation} 
 The above Hamiltonian is written in a multi-rotating frame defined by the unitary operator $U=\exp[-iVt]$, where $V=E_1\sigma_{11}+(E_1+\omega_p-\omega_s)\sigma_{22} + (E_1+\omega_p)(\sigma_{33}+\sigma_{44})$, such that after neglecting the oscillating terms the new Hamiltonian reads
 \begin{align}
 \tilde{H} &= \delta\sigma_{22} + \Delta_3\sigma_{33} + \Delta_4\sigma_{44} \nonumber \\ &+\frac{\Omega_p(t)}{2}\left(\cos(\theta)\sigma_{13}+\sin(\theta)\sigma_{14}\right)\nonumber \\
 &+\frac{\Omega_s(t)}{2}\left( \sin(\theta)\sigma_{23} - \cos(\theta)\sigma_{24} \right) + h.c.,
 \end{align}
with $\delta=\Delta_3-\Delta_2$, $\Delta_2=(E_3-E_2-\omega_s)$, $\Delta_3=E_3-E_1-\omega_p$ and $\Delta_4=E_4-E_1-\omega_p$.

%Bibliografía
\providecommand{\noopsort}[1]{}\providecommand{\singleletter}[1]{#1}%

\end{document}